# Errors in equations for galaxy rotation speeds

Kenneth F Nicholson,  Caltech Alumni
knchlsn@alumni.caltech.edu
nd364m@aol.com

**Abstract**

Shown are the errors and difficulties of the equations used for galaxy rotation speeds in the book "Galactic Dynamics" (Binney and Tremaine).  A usable and accurate set of equations is then presented.  The new equations allow easy determination of galaxy mass distribution from the rotation profile with no need for dark matter or any knowledge of galaxy surface light.

**Introduction**

Nearly everyone trying to find mass distribution in galaxies now assumes an exponential SMD distribution based on the surface light of a galaxy, then fills in the "missing mass" with dark-matter spherical shells to make the rotation speeds match experiment.  However, this "method" makes anyone with a little math background cringe, since it is such a misapplication of Newton's law.  So it is interesting to examine the reasons for this strange action, by people who must have a good math background to be in this field.  A lot of it can be blamed on blindly following the lead of those before in the many references.

The beginners in this field had no use for the effects of light, "gas" pressure, magnetic effects, or dark matter.  They assumed that the gravity effects of the matter in the galaxy envelope were enough to cause the observed rotation speeds, and they were right.  Perhaps their biggest mistake was in trying to find analytic results, instead of just adding up the effects in a brute-force method.  The result was that their methods only gave reasonable results for special cases, while a good brute-force method can handle almost any arbitrary input.  Either approach should not be trusted until checked against theory, such as the results for a sphere, or a single ring if it is done right.  The inability to handle the rotation speeds caused by a single ring is perhaps the reason most of the analytic methods failed.

**Discussion**

The Book "Galactic Dynamics" by Binney and Tremaine (B&T) has become a standard text for this subject, the best I have ever read.  Unfortunately the equations shown there for the purpose of finding the rotation profile from mass distribution, or mass distribution from the rotation profile, are either wrong or unworkable for <u>arbitrary</u> inputs, and this fact has led to the use of some bizarre methods.

The first of  their equations is 2-146, that attempts to solve the forward problem, ie find the rotation profile from a given arbitrary distribution of SMD.  It assumes a thin disk representing all the matter in a galaxy.  No dark matter spheres are used or implied, and the light/mass ratio is not mentioned.  If they could have gotten the correct result for a single ring, they could have gotten it right for the whole galaxy, but they didn't.  The upper limit on the integral can be changed to rmax without changing the result.

No method is proposed to solve the reverse problem  (mass distribution from rotation speeds) using this equation, so that important part is left out completely.  The equation is:

$$v^2(R,z) = \frac{G}{\sqrt{R}} \int_0^\infty \left[ K(k) - \frac{1}{4}\left(\frac{k}{1-k^2}\right) \times \left(\frac{r}{R} - \frac{R}{r} + \frac{z^2}{Rr}\right) E(k) \right] k \Sigma(r) \sqrt{r} \, dr \qquad \text{B \& T}(2-146)$$

$$\text{where } k^2 = \frac{4Rr}{(R+r)^2 + z^2}$$

$\Sigma(r) =$ SMD of ring at radius r

$(R, z) =$ location of test mass

As the authors remark, the integrand becomes infinite for k=1, where the test mass is at the same radius as one of the rings, and z=0. The infinities can be avoided by setting z at some other value than zero. But because thickness is important, it is not possible to choose a value for z to get the correct effect for all radii. This equation is now seldom used. However this form for the integral is sound, ie integrating the effects of each ring, of arbitrary SMD, on a test mass at a given location. For a single ring, thin in the radial direction but with thickness (say 5%) in the z direction, the acceleration of a test mass passing through it must stay finite and smoothly go through zero as it changes signs. If that could be arranged, this form might be be used . The functions K(k) and E(k) can't do this and should be avoided because of computer table look-up complexity, as well as the infinities.

Next is a set of two equations that suggest the SMD distributions can be arbitrary, but it is not so. Solutions are available only for a few special cases, and a thin disk is assumed. Another problem with these is that the upper limit cannot be finite. The equations are:

$$S(k) = -2\pi G \int_0^\infty J_0(kR) \Sigma(R) R \, dR, \qquad \text{B \& T}(2-158)$$

$$v^2(R) = -R \int_0^\infty S(k) J_1(kR) k \, dk, \qquad \text{B \& T}(2-160)$$

Two special cases result in solutions for v. The first is Mestel's disk:

$$\text{for } \Sigma = \frac{\Sigma_0 R_0}{R}, \quad v^2(R) = 2\pi G \Sigma_0 R_0 = \frac{GM(R)}{R}, \text{ constant} \qquad \text{B \& T}(2-165a)$$

where $M(R) = 2\pi \Sigma_0 R_0 R =$ mass inside R

So v is constant for all radii, and the authors say the mass outside R has no effect. Actually the mass outside R is infinite, and has a profound effect. The authors have assumed that the mass inside R acts like a sphere, but since it is in the form of a thin disk the gravity effects are much stronger on a test mass at R, and the mass outside R acts to keep the speed constant. When the disk has an outer rim (Rmax) and the mass outside Rmax is gone, the true rotation speeds build up toward the rim with a maximum at the rim, an effect that shows up in experimental data.

Many galaxies have nearly constant speeds to large radii, and for that part this equation (B&T 2-165a) looks good. To see how good it is, the results for this distribution are computed using the methods of Nicholson (2003), and are shown in figure 1. The constant speed part shown is the result of B&T2-165a, which is correct if the max radius is infinite. For this case the total mass would be high by a factor of 2.25 using the finite rim radius and a measured rim speed. This SMD distribution could also be claimed as being correct for the galaxy plane (instead of the exponential shown next). Then, trying to fudge the answers, sphere effects could be added with negative mass (wow!), or positive mass could be added outside the rim. Both methods would of course be wrong. No doubt the negative mass solution would make good headlines.

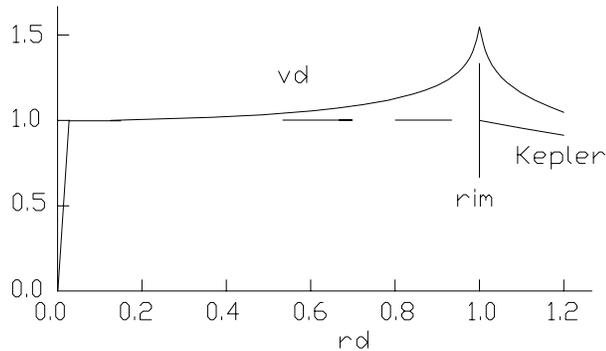

Figure 1. Dimensionless rotation speeds for .01 thick Mestel's disk, finite radius

The second special case of a solution based on B&T(2-158, 160) is nearly correct, except for the thickness effects, and the effects of removing the mass outside Rmax. It's the exponential distribution.

$$\Sigma = \Sigma_0 e^{-R/R_d}, \quad v^2(R) = 4\pi G \Sigma_0 R_d \, y^2 [I_0(y)K_0(y) - I_1(y)K_1(y)] \qquad \text{B \& T}(2-169)$$

where $y \equiv \dfrac{R}{2R_d}$

and $M(R) = 2\pi \Sigma_0 R_d^2 \left[ 1 - \exp(-R/R_d)\left(1 + \dfrac{R}{R_d}\right) \right]$

Note that $R_d$ is merely a shaping constant and not Rmax.

This equation, B&T(2-169) has been the equation of choice for nearly everyone using dark matter. The two constants, $\Sigma_0$ and $R_d$, cannot be found from the rotation speeds, so it is assumed that in the galaxy disk itself, the SMD is proportional to galaxy surface starlight (the mass/light ratio), and that since the surface starlight falls off about exponentially, the SMD must also. So this equation is used for the disk, and when that turns out to result in low values of rotation speeds, the difference is made up (fudged) using dark-matter spheres with positive mass. Figure 2 compares B&T(2-169) with Nicholson's method (solid line). Except at the rim the two are very close. At the rim the cross showing B&T(2-129) is slightly low because it assumes matter outside Rmax goes to infinity. That matter is now finite unlike that with Mestel's disk.

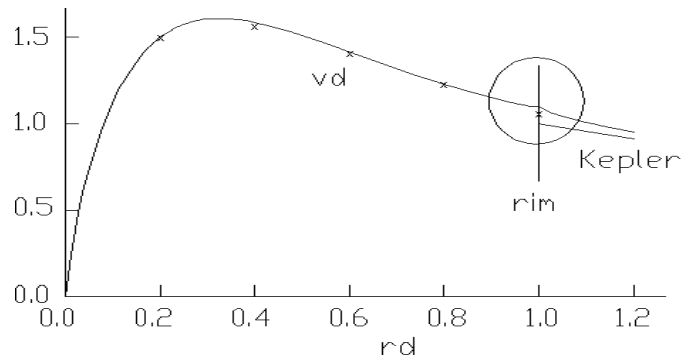

Figure 2. Dimensionless rotation speeds for $\Sigma = \Sigma_0 e^{-R/R_d}$, .01 thick

Now a lot of people have an almost religious faith in this equation and that procedure, but let's face it, adding the spheres with positive mass is no more correct than adding spheres with negative mass for Mestel's disk. Science is not a democratic procedure. Fifty references using dark matter don't make it right. The ApJ has accepted hundreds of papers based on this concept and never bothered to check the assumptions behind it. (The ApJ has also refused two of my papers). There are at least seven assumptions required for this procedure but they are never mentioned. Some are false.

1. SMD is always proportional to galaxy surface light.
    Partly true. However galaxies can exist without stars, and direct proportionality has not been proved.
2. All galaxy disks can be considered as zero thickness.
    False. Thickness has a significant effect, large for bulbous centers.
3. All galaxy disks have an exponential SMD distribution in radius.
    False. Very few galaxies have SMD's that are closely exponential.
4. If the assumed SMD distribution doesn't cause enough rotation speed, extra mass must be present.
    True
5. All extra mass must be added as spheres centered on the galaxy center.
    False. To satisfy the mathematics, the extra mass needed can be added almost anywhere near the galaxy, preferably in the galaxy envelope, where real matter is known to exist.
6. Light and gas pressures have negligible effect on rotation speeds.
    True, at least for heavy (non gas) objects.
7. Newton's law applies unchanged throughout the region of interest.
    True until proven otherwise

Another equation that is seldom used is B&T(2-174), intended to do the reverse problem, finding mass distribution from rotation speeds. The authors admit the derivatives in it can cause noise, and thus trouble, but there is a more serious problem than that. The equation has to be totally wrong because it fails completely for galaxies with long stretches of constant rotation speeds. It is:

$$\Sigma(r) = \frac{1}{2\pi G} \left[ \frac{1}{r} \int_0^r \frac{dv^2}{dR} K\left(\frac{R}{r}\right) dR + \int_r^\infty \frac{dv^2}{dR} K\left(\frac{r}{R}\right) \frac{dR}{R} \right], \qquad \text{B \& T}(2-174)$$

where r = radius of ring with SMD of $\Sigma$

    R = radius of test mass

    v = measured speed of teat mass at R

To this point in their book it is clear that B&T had no intention of using dark-matter spheres, there was no extra mass implied by gravity effects, and there was no need for a knowledge of light intensity. Only much later in their book did they succumb to these concepts. On page 599 they draw the wrong conclusion:

"There is no well established example of a Keplerian region in any galaxy rotation curve, even those that extend to radii large enough to contain all the galaxy's light. Consequently, *there is no spiral galaxy with a well determined total mass.* (italics are theirs)

"The simplest interpretation of these results is that other spiral galaxies, like our own, possess massive dark halos that extend to larger radii than the optical disks, a conclusion first stated by Freeman (1970)."

On page 602 B&T show a plot that demonstrates the dark-halo procedure, their figure 10-2, and their text implies that is the only way to do it.

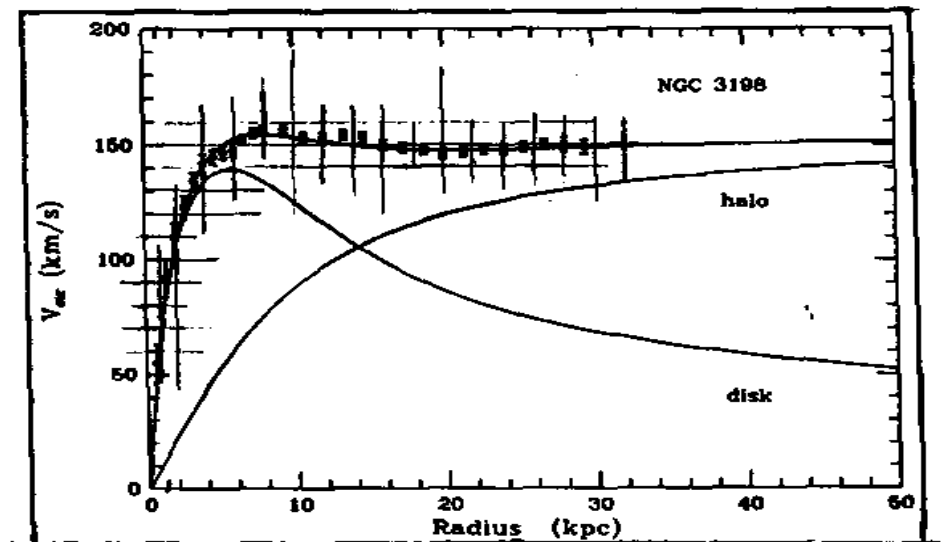

B&T Figure 10-2. The dark-halo method of fudging the rotation speeds,
    Van Albeda et al. (squares of speeds are added)

But of course there is a better way, shown in the next section. It can determine mass distribution and total mass quite well, in spite of B&T's conclusion above.

**A method that works**

As mentioned above, a usable method for the forward problem must be able to handle a single ring correctly, and account for the effects of thickness. The effects of an "atmosphere" (gas and dust extending above and below the large bodies) can be handled by using an equivalent thickness of constant density. Then the effects of all rings making up the galaxy can be added up to find the acceleration and speed of a test mass at any radius. The derivation and proofs are given by Nicholson (2003). In digital form, the equation for the forward problem is:

$$v^2 = (\text{pytks})^2 \times \text{rt} \times \sum_{1}^{Nr} 2 \sum_{1}^{180} \frac{G \ ddm}{\sqrt{c^2 + (h/2)^2}} \times \frac{(rt - rr \cos(th))}{c^2} \qquad (1)$$

where $c^2 = (rr \sin(th))^2 + (rt - rr \cos(th))^2$, $pc^2$, note that c is never zero **

ddm = mass of the fundamental segment, rho h r dth dr , msuns
dth = 1 degree
G = gravitational constant, 4.498E-15 $pc^3$ / (msuns / $yr^2$ )
h = galaxy equivalent thickness at radius r, pcs
Nr = number of rings
pytks = 9.778E5 (kms/sec)/(pcs/yr)
r = radius to centerline of ring, pcs
rho = equivalent density, msuns/$pc^3$
rm = radius to outer edge of ring, pcs
rr = radius to rod used to represent fundamental segment mass, pcs
= rm -dr / 2 × (rm - 0.575 dr) / (rm - dr / 2) *
rt = radius to test mass, pcs
th = (i-1/2) dth, for i = 1 to 180 , degs
v = orbital speed of test mass at radius rt, kms/sec

* The position of the rod in the fundamental segment was moved slightly from the centroid toward the mid radius of the segment, based on trials to return a given mass distribution from a computed rotation profile.
** Since the rods are at the midangle of each segment, ie at th = 0.5 deg in the first and 179.5 degs in the last segments, c (the distance from test mass to rod) is never zero.

The method for the reverse problem finds the mass distribution that causes the measured data, by repeated trials and corrections using the forward method. The computing procedure used is:

1. input galaxy dimensions, measured speeds at the outer edges of each ring, and arbitrary starting densities for each ring (usually just one density for all rings)

2. compute rotation speeds at the outer edges of each ring using (1)

3. use speed errors to correct densities of the rings (i = 1 to Nr)

$\quad$ errv = (vm–v) / vmax , f = 0.75 errv , if all errv < 1E–6 then go to 5 $\qquad (2)$

$\quad$ limit abs(f) < 0.5 , rho(j) = (1+f) rho(j–1) $\quad$ for each cycle j

$\quad$ where vm = measured speed and vmax = maximum measured speed

4. go to 2 for the next cycle

5. make results dimensionless for plots, print or plot results and quit

Results include total mass, volume, average density, average SMD, Kepler rim speed,, maximum computed speed, and the plotted data: md, rhod, rSMDd, vd.

**An example**

The example chosen is NGC 3198 as in B&T figure 10-2. Measured data are the midpoints of those on the plot. Equivalent thickness was assumed to be that of the Milky Way as defined in arXiv 0309762. The density in the z direction can be found after finishing the problem, from the data in the same reference.

After computing is done in dimensioned form, all output data are made dimensionless by dividing with normalizing parameters:

$accd = acc / acckep$, where $acckep = GM / rmax^2$, the Kepler acceleration at the rim
$md = m(r) / M$, where $m$ = mass inside r, and $M$ = total galaxy mass
$hd, rd, rrd, rtd = (h, r, rr, rt) / rmax$, where $rmax$ = galaxy rim radius
$rhod = rho / rhoav$, where $rhoav = M / $ (total volume using equivalent segments)
$SMDd = SMD / SMDav$, where $SMDav = M / (\pi\, rmax^2)$
$rSMDd = rd \times SMDd$
$vd, vmd = (v, vm) / vkrim$, where $vkrim = sqr(GM / rmax)$, the Kepler speed at the rim

At problem end the computed speeds from the mass distribution are essentially identical with the inputs. The output of this method is as good as the galaxy dimensions and measured speeds. By putting the results in dimensionless form, the similarities and differences between the dynamics of all sizes of galaxies

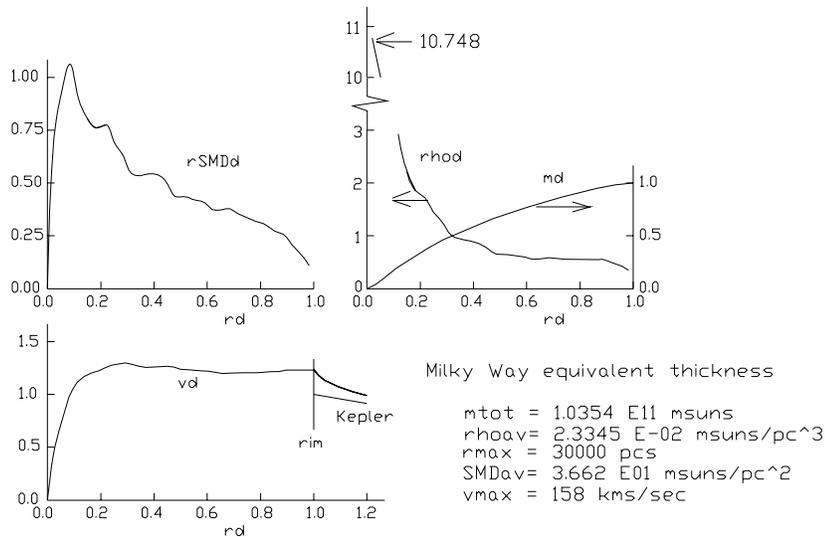

Figure 3. Complete results for NGC 3198

**References**
Binney, J. and Tremaine, S. (1987), *Galactic Dynamics*, Princeton University Press
Nicholson, K. F. (2003), lanl.arXiv.org, astro-ph/0309762, "Galaxy statics without dark matter"
Nicholson, K. F. (2003), lanl.arXiv.org, astro-ph/0303135, "Galaxy mass distributions from rotation speeds by closed-loop convergence"
Nicholson, K. F. (2001), lanl.arXiv.org, astro-ph/0101401, "Galaxy mass distributions for some extreme orbital –speed profiles"
Nicholson, K. F. (2000), lanl.arXiv.org, astro-ph/0006330, "Disk-galaxy density distributions using Newton's law, version 1.1"
Nicholson, K. F. (2000), lanl.arXiv.org, astro-ph/0006140, "Disk-galaxy density distributions using Newton's law"